\documentclass {aa}
\usepackage{graphics}
\usepackage{psfig}
\input epsf.sty
\begin{document}

\newcommand{\eff}{\mbox{Effelsberg}}
\newcommand{\gb}{\mbox{Green Bank}}
\newcommand{\nan}{\mbox{Nan\c{c}ay}}
\newcommand{\as}[2]{$#1''\,\hspace{-1.7mm}.\hspace{.1mm}#2$}
\newcommand{\am}[2]{$#1'\,\hspace{-1.7mm}.\hspace{.0mm}#2$}
\newcommand{\dgr}{\mbox{$^\circ$}}   
\newcommand{\E}[1]{\mbox{${}\,10^{#1}{}$}}
\newcommand{\etal}{et al.}
\newcommand{\grd}[2]{\mbox{#1\fdg #2}}
\newcommand{\gsim}{\stackrel{>}{_{\sim}}}
\newcommand{\lsim}{\stackrel{<}{_{\sim}}}
\newcommand{\Ha}{\mbox{H$\alpha$}}
\newcommand{\HI}{\mbox{H\,{\sc i}}}
\newcommand{\HIbf}{\mbox{H\hspace{0.155 em}{\footnotesize \bf I}}}
\newcommand{\HIit}{\mbox{H\hspace{0.155 em}{\scriptsize \it I}}}
\newcommand{\HIsl}{\mbox{H\hspace{0.155 em}{\footnotesize \sl I}}}
\newcommand{\HIss}{\mbox{H\,{\sc i}}}
\newcommand{\HII}{\mbox{H\,{\sc ii}}}
\newcommand{\jks}{\mbox{Jy$\cdot$km s$^{-1}$}}
\newcommand{\kms}{\mbox{ km\,s$^{-1}$}}
\newcommand{\kmsMpc}{\mbox{ km\,s$^{-1}$\,Mpc$^{-1}$}}
\newcommand{\LB}{\mbox{$L_{B}$}}
\newcommand{\LBnul}{\mbox{$L_{B}^0$}}
\newcommand{\LBsun}{\mbox{$L_{\odot,B}$}}
\newcommand{\Lsun}{\mbox{$L_{\odot}$}}
\newcommand{\LsunMsun}{\mbox{$L_{\odot}$/${M}_{\odot}$}}
\newcommand{\LFIR}{\mbox{$L_{FIR}$}}
\newcommand{\LFIRLB}{\mbox{$L_{FIR}$/$L_{B}$}}
\newcommand{\LFIRLBnul}{\mbox{$L_{FIR}$/$L_{B}^0$}}
\newcommand{\LFIRLsun}{\mbox{$L_{FIR}$/$L_{\odot,Bol}$}}
\newcommand{\MHI}{\mbox{${M}_{HI}$}}
\newcommand{\mhi}{\mbox{${M}_{HI}$}}
\newcommand{\MHILB}{\mbox{${M}_{HI}$/$L_{B}$}}
\newcommand{\MHILBnul}{\mbox{${M}_{HI}$/$L_{B}^0$}}
\newcommand{\Msun}{\mbox{${M}_\odot$}}
\newcommand{\msun}{\mbox{${M}_\odot$}}
\newcommand{\MsunLsun}{\mbox{${M}_{\odot}$/$L_{\odot,Bol}$}}
\newcommand{\MsunLBsun}{\mbox{${M}_{\odot}$/$L_{\odot,B}$}}
\newcommand{\MT}{\mbox{${M}_{ T}$}}
\newcommand{\MTLBnul}{\mbox{${M}_{T}$/$L_{B}^0$}}
\newcommand{\MTLBsun}{\mbox{${M}_{T}$/$L_{\odot,B}$}}
\newcommand{\mi}{\mbox{$\mu$m}}
\newcommand{\NH}{\mbox{N$_{HI}$}}
\newcommand{\OIII}{\mbox{[O\,{\sc iii}]}}
\newcommand{\s}{\mbox{$\sigma$}}
\newcommand{\Tb}{\mbox{$T_{b}$}}
\newcommand{\tis}[2]{$#1^{s}\,\hspace{-1.7mm}.\hspace{.1mm}#2$}
\newcommand{\vhel}{\mbox{$V_{hel}$}}
\newcommand{\vrot}{\mbox{$v_{rot}$}}
\def\la{\mathrel{\hbox{\rlap{\hbox{\lower4pt\hbox{$\sim$}}}\hbox{$<$}}}}
\def\ga{\mathrel{\hbox{\rlap{\hbox{\lower4pt\hbox{$\sim$}}}\hbox{$>$}}}}

\thesaurus{03(11.04.1;   % Galaxies: distances and redshifts
              11.07.1;   % Galaxies: general
              11.09.4;   % Galaxies: ISM
              13.19.1)}  % Radio lines: galaxies

\title{H\,{\large \bf I} observations of giant low surface brightness galaxies}

\author{L.D. Matthews\,\inst{1}, W. van Driel\,\inst{2} 
   \and D. Monnier-Ragaigne\,\inst{2,3}}

\offprints{L.D. Matthews, e-mail: lmatthew@nrao.edu}

\institute{National Radio Astronomy Observatory, 520 Edgemont Road,
 Charlottesville, VA 22903, U.S.A.
\and
Unit\'e Scientifique \nan, USR CNRS B704,
 Observatoire de Paris, 5 place Jules Janssen, F-92195 Meudon, France
\and
DAEC, UMR CNRS 8631
Observatoire de Paris, 5 place Jules Janssen, F-92195 Meudon, France}

\date
{\it Accepted for Astronomy and Astrophysics Supplements}

%Received  ; accepted }
\maketitle

\markboth{{L.D. Matthews et al.: \HI\ observations of Giant LSB galaxies}}{}

\begin{abstract}
We have used the \nan\ Radio Telescope to obtain new global \HI\ data
for 16 giant low surface brightness (LSB) galaxies.  
Our targets have optical luminosities and disk scale lengths at the 
high end for spiral galaxies ($L_{B}\sim 10^{10}$~\Lsun\ and
$h_{r}\ga$6~kpc for $H_{0}$=75~\kms~Mpc$^{-1}$), 
but they have diffuse stellar disks, with mean
disk surface brightnesses $\ga$1 magnitude 
fainter than normal giant spirals. Thirteen of the galaxies 
previously had  been detected in 
\HI\ by other workers, but the published \HI\ observations were either
confused, resolved by the telescope beam, of low signal-to-noise, or 
showed significant discrepancies between
different authors. For the other 3 galaxies, no \HI\ data 
were previously available.
Several of the galaxies were resolved by the \nan\
\am{3}{6} E-W beam, so global parameters were derived from
multiple-point mapping observations. Typical \HI\ masses for 
our sample are $\ga 10^{10}$~\msun, with $M_{HI}/L_{B}=$0.3-1.7 (in
solar units). All of the observed galaxies have 
published optical surface photometry, and we have compiled key optical
measurements for these objects from the literature. 
We frequently find significant variations among
physical parameters of giant LSB galaxies reported by
various workers.

\end{abstract}

\section{Introduction}
\subsection{The discovery of giant low surface brightness galaxies}
In 1987, Bothun et al. reported the serendipitous discovery of 
the extraordinarily large low surface brightness (LSB) 
galaxy now known as Malin-1. In spite of 
having a projected $B$-band central surface brightness of only
26.5~mag arcsec$^{-2}$,  Malin-1 is the largest spiral galaxy known, 
with a disk scale length of 73~kpc (assuming $H_0$=75 \kms\ Mpc$^{-1}$), and an
exceptionally high \HI\ mass ($\sim 10^{11}$\Msun) 
and optical luminosity ($M_{B}=-$23.1). 
Subsequently, through systematic 
searches of photographic survey plates, other galaxies with 
similar (albeit slightly less
extreme) properties to Malin-1
have been uncovered (Bothun et al. 1990; 
Sprayberry et al. 1993; Sprayberry et al. 1995b). We hereafter refer
to these as ``LSB Giants''. A handful of
LSB Giants are also found
in the UGC (Nilson 1973), NGC (Dreyer 1953), and ESO (Lauberts \&
Valentijn 1989) catalogues (see Gallagher \& Bushouse 1983;
Impey \& Bothun 1989;
Walsh et al. 1997; Pickering et al. 1997; Schombert 1998).
Nonetheless, while  recent photographic and CCD surveys have 
uncovered large numbers of new small and medium-sized, moderate-to-low 
surface brightness  spiral galaxies (e.g., Schombert et al. 1992;
Impey et al. 1996; O'Neil et al. 1997), 
LSB Giants have remained relatively rare. Having
faint, diffuse disks, but 
sizes, \HI\ masses, and luminosities at the high end for disk galaxies,
the LSB Giants occupy a unique realm of physical
parameters space and may share
evolutionary histories distinct from other LSB galaxies 
(e.g., Hoffman et al. 1992).

Since  a continuum of values exists for galaxy properties such as 
surface brightness, luminosity, and scale length, 
Sprayberry et al. (1995b) proposed to define LSB Giants as those 
objects meeting a ``diffuseness 
index'' criterion: $\mu_{B}(0)+5{\rm
log}(h_{r})>27.6$\footnote{Sprayberry 
et al. (1995b) formulated
their diffuseness index based on a distance scale with 
$H_{0}$=100~\kms~Mpc$^{-1}$. Here we have rewritten
the diffuseness criterion for $H_{0}$=75~\kms~Mpc$^{-1}$, which is
assumed throughout this work.}, where 
$\mu_{B}(0)$ is the extrapolated, deprojected
$B$-band disk central surface brightness in 
magnitudes arcsec$^{-2}$, and $h_{r}$ is the 
disk scale length in kpc. Among seven LSB 
Giants described by Sprayberry et al. (1995b), mean properties 
include: $<B-V>=0.73\pm0.05$, $<\mu_{B}(0)>=23.23\pm0.19$ mags
arcsec$^{-2}$, and $<h_{r}>=13.0$~kpc. 
The colors of these LSB 
Giants are thus comparable to those of normal spirals (Sprayberry 
et al. 1995b), 
but are redder than typical colors of many small and 
moderate-sized LSB disks (e.g., McGaugh \& Bothun 1994; Matthews \& 
Gallagher 1997; de Blok et al. 1996; Beijersbergen et al. 1999). 
In addition, the LSB Giants are distinct from other more common
LSB spirals in that they often have a 
significant bulge component (e.g., Gallagher \& Bushouse 1983;
Knezek 1993,1998), and
frequently their centers harbor an
active nucleus  (e.g., Schombert 1998).

The origin and evolutionary histories of LSB Giant 
galaxies are still enigmatic. Hoffman et al. (1992) have proposed a 
formation scenario whereby these systems form in very low density 
regions from 
rare, 3$\sigma$ density fluctuations. They predict 
these galaxies should exhibit quiescent, unevolved, gas-rich disks, with 
rotation curves that flatten near $V_{max}\sim$300~\kms. Knezek (1993)
has suggested an alternative scenario, based on Kormendy (1989),
whereby LSB Giants may have
dissipatively formed from massive, metal-poor dark matter halos.

\subsection{The need for new \HIit\ observations}
Testing formation and evolution scenarios for LSB Giants 
requires an accurate knowledge of the  neutral gas 
properties and linewidths of these galaxies. And only by 
combining such measures 
with optical data can we begin to build a picture of the 
star-formation histories of these systems and their relationship to
other types of LSB galaxies.

Other motivations also exist for improved \HI\ observations.
Hoffman et al. (1992) have argued that for the enormous disks of LSB Giants
to remain quiescent over a Hubble time, they must be very isolated. 
Yet studies 
hint that LSB Giants are in fact less isolated than other 
LSB spirals, although redshift surveys in the vicinities of these 
objects are still incomplete (Sprayberry et al. 1995b). 
Pointed \HI\ observations in the vicinity of LSB Giants can thus reveal 
if these galaxies have any yet-undiscovered gas-rich neighbors. 

Another important use of \HI\ data is for exploring the Tully-Fisher
(TF) relation 
for giant LSB spirals.
Sprayberry et al. (1995a) have shown that at least two
of the presently known LSB Giant galaxies are extreme outliers from
the TF relation 
defined by normal galaxies. This is unlike the bulk of 
moderate-sized, moderate luminosity 
LSB galaxies, which tend to follow TF
(Sprayberry et al. 1995a; Zwaan et al. 1995; Verheijen 1997). 
It is of considerable interest
therefore to assess from a larger sample whether LSB Giants deviate
systematically from the TF relation. 

While previous
\HI\ observations 
have established that LSB Giants are in general very gas-rich
(\MHI$\ga10^{10}$\msun; e.g., Sprayberry et al. 1995b; Walsh et al. 1997;
Pickering et al. 1997,1999), unfortunately
existing \HI\ data for many LSB Giants are of dubious quality (i.e., 
the galaxy was confused or resolved by the telescope beam, 
the spectra are of low
signal-to-noise, or measurements from different workers are highly 
discrepant; see also Table~5 and Section~\ref{disc}). 

For example, based on Arecibo 21-cm observations of 3 objects,
Sprayberry et al. (1995b)  suggested that peculiar, 
asymmetric \HI\ profiles may be commonplace for LSB
Giants. However, since the \HI\ extents of these galaxies were
expected to be comparable to the size of the telescope beam, it is
important to verify that these ``peculiar'' spectra do not
result from some combination of source resolution and telescope
mispointing. In addition, independent checks on derived \HI\
parameters are valuable since it is more difficult to accurately
measure integrated
fluxes and linewidths when the global \HI\ profiles are quite
broad compared to the bandwidth used, and when sources are at large
recessional velocities ($V_{r}\ga$ 15\,000~\kms), 
where flux calibration can become
increasingly uncertain. Finally, there are still a handful of known
LSB Giants for which no \HI\ data have previously been obtained. 

Quality single-dish \HI\ spectra for LSB Giants are also
a useful precursor and complement to \HI\ aperture synthesis studies 
of these galaxies (e.g., Walsh et al. 1997; Pickering et
al. 1997,1999). \HI\ mapping is of course a critical part of
understanding the dynamics and gas distributions for these galaxies, 
but since many of the known examples of LSB Giants are 
rather distant ($V_{r}>$10\,000~\kms) and of relatively modest optical angular 
size ($D_{25}<2'$), such observations are challenging and 
benefit from careful planning based on prior \HI\ measurements. 
Moreover, because LSB Giants are generally expected to have  disks
with relatively low \HI\ surface densities (e.g., Pickering et al. 1997),
 diffuse emission can be missed in
aperture synthesis measurements, and
total flux  and maximum rotational velocity measures from \HI\ pencil 
beam observations serve as an important check.

Based on the above motivations, we have used the \nan\ Radio Telescope
to obtain new global 
observations of a sample of 
16 LSB Giant galaxies for which existing global \HI\ measurements
were incomplete, required reconfirmation, or were 
nonexistent. We measure integrated \HI\ line
fluxes, linewidths, and recessional
velocities,
and  attempt to
clear up conundrums surrounding several of these objects in the literature.

\section{Sample selection}
Our targets for the present program were 16 LSB Giant galaxies 
meeting the ``diffuseness'' criterion stated in Section~1. The objects 
were culled from  Knezek (1993), Sprayberry et al. (1995b), and 
Pickering et al. (1997), and all have available  optical surface 
photometry. Thirteen
of our targets have previously been observed in 
\HI\ by other workers, but the published \HI\ parameters were
uncertain or incomplete (see below).
For the other 3 galaxies, no \HI\ data were previously available. 
Some basic parameters for our target galaxies are summarized in Table~1.
Coordinates, $B$ magnitudes, 
angular sizes, axial ratios, and radial 
velocities were culled from the NED database. Position angles and 
scatter in the
published radial velocity values [$\sigma(V)$] were taken from the LEDA
database, and inclinations were obtained from the various references given in 
Table~6 (discussed below). 
Inclinations and position angles marked with a colon (:) were
estimated by us from inspection of the Digitized Sky Survey.

% -----------------------------------------------------------------
%
% --- Table 1: Basic data 
%
\begin{table*}
\caption{Basic parameters for target galaxies compiled from NED and LEDA}
{\footnotesize
\begin{tabular}{lccllrrrrrlrr}
\hline
\vspace{-2 mm} \\
Ident. & R.A. & Dec & \multicolumn{2}{c}{Classification} & $m_{b}$ 
& $D_{25}$ & $b/a$ & $i$ & $PA$ & $V_{r}$ & $\sigma(V)$ \\ 
 & \multicolumn{2}{c}{(1950.0)} & {\scriptsize LEDA} & {\scriptsize NED} & mag 
  & $'$ & & $^{\circ}$ &  $^{\circ}$ & km/s & km/s \\
\vspace{-2 mm} \\
\hline
\vspace{-2 mm} \\
UGC 568    & 00 52 35.5 & -01 19 01 & S?  & Sd       & 15.08 & 1.17 &
 0.70 & 40 & & 13197 & 204  \\
UGC 1752   & 02 13 30.0 &  24 40 00 & Sc  & SA(s)cd  & 16.5  & 1.55 &
 1.00 & 0: &0: & 17847 &  60  \\
PGC 135657 & 02 37 38.7 & -01 59 18 & Sb  & Sc       & 15.47  & 0.78 &
      & 38 & 63 & 12701 &  60  \\
UGC 2936   & 04 00 12.6 &  01 49 36 & SBc & SB(s)d   & 15.00  & 2.51 &
 0.27 &78 & 30 & 3817 & 125  \\
UGC 3140   & 04 40 20.1 &  00 31 35 & Sbc & SA(rs)c: & 13.30 & 1.78 &
 0.91 & 0: & 175 & 4633 &  60  \\
NGC 2770   & 09 06 29.8 &  33 19 53 & Sc  & SA(s)c:  & 12.77 & 3.66 & 
 0.30 &  79 & 148 & 1947 &  63  \\
PGC 135754 & 10 34 52.9 &  02 20 57 & Sc  & Sc       & 16.34  &      &
      & 49 & & 21335 &  60  \\
F568-6     & 10 37 08.5 &  21 06 24 & Scd & Sd/p     & 15.28  & 1.67 &
 0.79 & 38 & & 13830      &      \\
UGC 6614   & 11 36 39.1 &  17 25 14 & Sa  &(R)SA(r)a:& 14.37 & 2.02 &
 0.86 &  35 & 0: & 6351 &  60  \\
Malin-1    & 12 34 27.3 &  14 36 15 &     & S        & 16.20  & 0.20 &
 0.93 & 45 & 45 & 24750      &      \\
PGC 45080  & 13 00 42.6 &  01 44 12 & Sc  & Sc       & 15.61  & 0.79 &
 0.22 & 72 & 84 & 12264 & 295  \\
UGC 9024   & 14 04 21.0 &  22 18 28 & Sa  & S:       &  16.0     & 2.09 &
 0.95 & 37 & 0: & 2323      &      \\
F530-1     & 21 05 21.5 &  26 15 00 & Sc  & S(r)     & 16.31  & 0.47 &
 0.74 & & & 14340      &      \\
F533-3     & 22 14 53.3 &  24 57 48 & SBc & SBc(r)   & 15.43  & 0.88 &
 0.65 &65: &165: & 12669       &      \\
NGC 7589   & 23 15 41.9 & -00 00 45 & Sb  & SAB(rs)a:& 15.01  & 1.07 &
 0.62 &  58 & 60 & 8938 & 219  \\
PGC 71626  & 23 27 58.2 & -02 44 18 & SBc &SB(r)b pec& 15.14  & 1.78 &
 0.59 &  61 & 65 & 9520 & 247  \\
\vspace{-2 mm} \\
\hline
\end{tabular}
}
\end{table*}
% -----------------------------------------------------------------
%
% --- Table 2
%
\begin{table*}%
\caption{New measured global \HI\ parameters}
\begin{footnotesize}
\begin{tabular}{lcccccccccl}
\hline
\vspace{-2 mm} \\
 Galaxy Name  & rms  & $F_{max}$ & $W_{20}$  & $W_{50}$  &
  $V_{HI}$ & $\sigma(V)$ & $S_{HI}$ & $\sigma(S)$ & S/N & Notes \\
\vspace{-2 mm} \\
  & mJy & mJy & \kms\ &  \kms\ & \kms\ & \kms\ & \jks\ & \jks\ &  &  \\

~~~~~~~~(1) & (2) & (3) & (4) & (5) & (6) & (7) & (8) & (9) & (10) & (11) \\
\vspace{-2 mm} \\
\hline
\vspace{-2 mm} \\
UGC~568 & 1.15 & 12.3 & 410: & 300: &  13420 & 10 & 2.28 & 0.28 & 10.7 & peculiar\\

UGC~1752* & 2.84 & 26.6 & 429 & 388 & 17861 & 7 & 5.13 & 0.70 & 9.4 &  \\

PGC~135657 & 2.71 & 23.8 & 134 & 100 & 13108 & 7 & 2.17 & 0.46 & 8.8 & blended?\\

UGC~2936* & 2.32 & 38.6 & 502 & 469 &  3811 & 4 & 11.23 & 0.72 & 16.7 & \\

UGC~3140* & 4.01 & 130.8 & 131 & 150 &  4623 & 2 & 13.41 & 0.74 & 32.6 &\\

NGC~2770* & 4.84 & 143.6 & 350 & 322 &  1947 & 2 &36.93 & 1.41 & 29.6 & \\

PGC~135754 & 1.65 & 5.5 & 222 & 197 &  21769 & 15 & 0.59 & 0.30 & 3.4 &  \\

F568-6* & 2.29 & 17.5 & 420 & 393 &  13829 & 7 & 4.20 & 0.64 & 7.6 & \\

UGC~6614* & 2.86 & 81.2 & 290 & 255 &  6351 & 2 & 15.94 & 0.72 & 28.4 & \\

Malin-1* & 2.02 & 9.4 & 341 & 293  & 24784 & 15 & 1.80 & 0.50 & 4.7 & \\

PGC~45080 & 2.90 & 11.5 & 315 & 280 &  12339 & 15 & 1.57 & 0.61 & 4.0 &\\

UGC~9024* & 2.69 & 115.7 & 134 & 120 & 2321 & 1 & 11.41 & 0.48 & 42.9 &\\

F530-1 & 3.0 & 10.7 & 497 & 484 &  14335 & 10 & 2.07 & 0.75 & 3.6 & \\

F533-3* & 2.72 & 13.3 & 399 & 384 &  12670 & 8 & 2.46 & 0.66 & 4.9 & \\

NGC~7589 & 1.68 & 10.4 & 414: & 357: & 8866 & 12 & 2.96 & 0.51 & 6.2 & blended\\

PGC~71626* & 3.40 & 47.0 & 126 & 88 &  10016 & 5 & 3.94 & 0.56 & 13.8 & \\
\vspace{-2 mm} \\
\hline
\vspace{-2 mm} \\
\multicolumn{11}{l}{Notes: *\, indicates mapped galaxy (see Fig.~2 and Sect.~3.5).}
\end{tabular}
\end{footnotesize}

\end{table*}
%---------------------------------------------------------------------------------

\section{Observations} 
\subsection{Data acquisition}

The \nan\ decimentric radio 
telescope is a meridian transit-type instrument of the 
Kraus/Ohio State 
design, consisting of a fixed spherical mirror (300~m long and 35~m high),
a tiltable flat mirror (200$\times$40~m), and a focal carriage moving along a 
curved rail track. Due to an ongoing major renovation of the focal system, the
length of the focal track was reduced to $\sim$60~m during the period
of our observations, thus allowing
tracking of a source on the 
celestial equator for about 45 minutes. The effective collecting 
area of the \nan\ telescope
is roughly 7000~m$^{2}$ (equivalent to a 94-m diameter parabolic dish). 
Due to the elongated geometry of the mirrors, at 21-cm wavelength the \nan\
telescope has a 
half-power beam width of \am{3}{6}~E-W$\times$ 22$'$~N-S for the range of 
declinations covered in  this work (E. G\'erard, private comm.; see also 
Matthews \& van Driel 2000). Typical system temperatures were $\sim$40~K for 
our project.

The observations at \nan\ were made in the period June 1998 - October 1999, 
using a total of about 300 hours of telescope time.
We obtained our observations in total 
power (position-switching) mode using consecutive pairs 
of two-minute on- and two-minute 
off-source integrations. Off-source integrations were taken at 
approximately 20$'$~E of the target position.
The autocorrelator was divided into two pairs
of cross-polarized receiver banks, each with 512 channels and a 6.4~MHz 
bandpass. This yielded a channel spacing of 2.64~\kms, for an effective
velocity resolution of $\sim$3.3~\kms\ at 21-cm. The center 
frequencies of the two banks were tuned to the expected redshifted \HI\ 
frequency of the target based on values from the literature
(Table~1). Depending on the
signal strength, the spectra were smoothed 
to a channel separation of $\sim$7.9 or $\sim$13.2~\kms\
during the data reduction in order to increase signal-to-noise.  
Total integration times were up to 
12 hours per galaxy, depending on the strength of the source and
scheduling constraints.

In all cases, data were initially obtained with
the telescope pointed at the published
optical center of the galaxy. However, as shown by van der Hulst et al. (1993),
the \HI\ disks 
of large LSB spirals may frequently
extend to up to $\sim$2.5$\times$ their optical diameter.  
We therefore observed several of the targets, including the 8 galaxies
with $D_{25}\ga$\am{1}{3}
(i.e. one-third the \nan\ FWHP E-W beamwidth) at three 
or more spatial positions: one at the target's optical center, plus
additional pointings offset to the east or west by multiples of one-half
beamwidth (see Table~3, discussed below).
 Because of
the large N-S diameter of the \nan\
beam ($\ge$22$'$), these mapping observations were limited to pointings
along an E-W line. 
%
%----- Figure 1 ------------------------------------------------------
%
\begin{figure*}[!t]
%\vspace*{1cm}
\hfil \epsfxsize 14.5cm \epsfbox{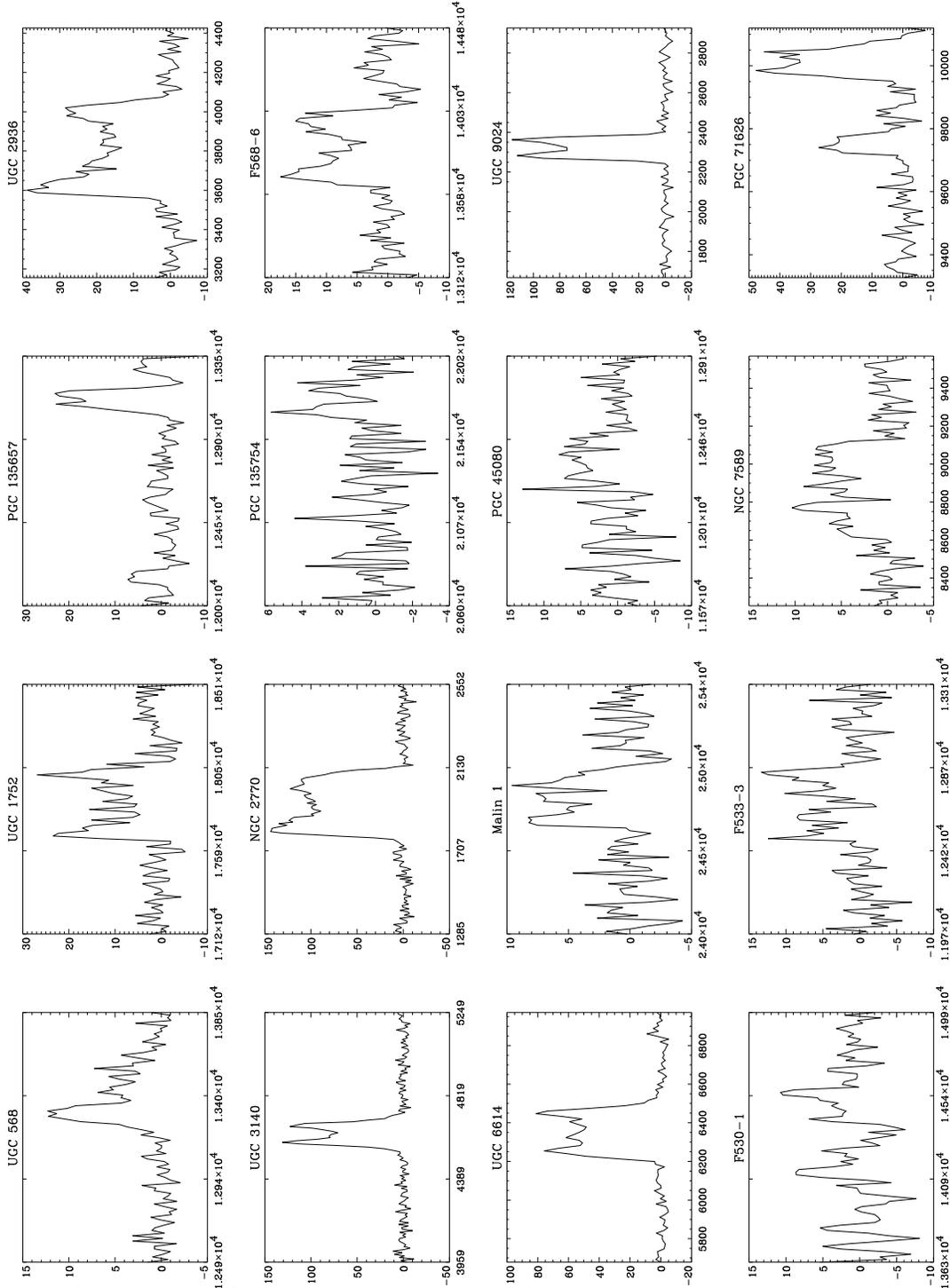}\hfil
\caption [] {\nan\ global \HI\ 21-cm line spectra of 16 Giant low
surface brightness galaxies. Axes are
flux density, in millijanskys, and radial velocity, in \kms, using the optical
convention. The spectra shown have been smoothed to velocity resolutions of
$\sim$8 or 13~\kms.}
\label{fig:spectra}\protect
\end{figure*}

%---------------------------------------------------------------------

\subsection{Calibration\protect\label{cal}}
Flux calibration (i.e.,$T_{sys}$-to-mJy conversion) 
at \nan\ is determined via regular measurements of a cold load
calibrator and periodic monitoring of strong continuum sources by 
the Nancay staff. Standard calibration procedures
include correction for
declination-dependent gain variations of the telescope (e.g., Fouqu\'e
et al. 1990). These techniques
typically yield an internal calibration accuracy of $\sim$15\% at frequencies
near 1420MHz.

%===========================================================================

%
% -----------------------------------------------------------------
%
% --- Table 3: HI mapping data
%
\begin{table}
\caption{Results from mapping observations}
{\scriptsize
\begin{tabular}{lcccc}
\hline
\vspace{-2 mm} \\
Ident. & Postion & rms & $S_{HI}$ & $V_{c}$ \\
  & & mJy & \jks & \kms \\
\vspace{-2 mm} \\
\hline
\vspace{-2 mm} \\
UGC~1752 & $2'$W & 3.00 & 2.42 & 17858 \\
         & C     & 2.45 & 4.72 & 17846 \\
         & $2'$E & 4.65 & 1.84$^{a}$ & 18028 \\
\\
UGC~2936 & 4$'$W & 2.55 & $\le$0.67 &  \\
         & 2$'$W &  3.55 & 5.99 &  3797 \\
         & C     & 2.76 & 9.83 &  3812 \\
         & 2$'$E & 2.56 & 4.29 &  3807 \\
         & 4$'$E & 2.31 & $\le$0.29 &  \\
\\
UGC~3140 & 6$'$W & 2.59 & 0.74 & 4596 \\
         & 4$'$W & 1.98 & 0.99 & 4662 \\
         & 2$'$W & 2.28 & 6.33 & 4618 \\
         & C     & 2.32 & 10.60 & 4626 \\
         & 2$'$E & 3.32 & 6.09 & 4628 \\
         & 4$'$E & 3.44 & 1.40 & 4632 \\
         & 6$'$E & 2.91 & $\le$0.27 & \\ 
\\
NGC~2770 & 6$'$W & 3.59 & 1.39 & 1908 \\
         & 4$'$W & 3.38 & 4.08 & 1966 \\
         & 2$'$W & 4.00 & 16.16 & 1950 \\
         & C     & 2.56 & 28.50 & 1948 \\
         & 2$'$E & 3.46 & 15.58 & 1930 \\
         & 4$'$E & 2.86 & 3.82 & 1922 \\
         & 6$'$E & 4.15 & 1.91 & 2022 \\
\\
F568-6   & 2$'$W & 2.88 & 2.27 & 13825 \\
         & C     & 3.36 & 4.56 & 13812 \\
         & 2$'$E & 2.94 & 1.20 & 13948 \\
\\
UGC~6614 & 6$'$W & 2.02 & $\le$0.15 & \\
         & 4$'$W & 2.95 & 1.73 & 6437 \\
         & 2$'$W & 3.06 & 10.28 & 6368 \\
         & C     & 2.00 & 10.39 & 6347 \\
         & 2$'$E & 3.04 & 5.99 & 6338 \\
         & 4$'$E & 2.09 & 1.36 & 6329 \\
         & 6$'$E & 1.79 & 0.80 & 6355 \\
\\
Malin~1  & 2$'$W & 1.37 & 0.89 & 24750 \\
         &  C    & 2.08 & 1.63 & 24774 \\
         & 2$'$E & 2.08 & 0.96 & 24794 \\
\\
UGC~9024 & 4$'$W & 2.63 & 1.08 & 1293 \\
         & 2$'$W & 2.76 & 6.04 & 2323 \\
         & C     & 3.30 & 9.98 & 2323 \\
	 & 2$'$E & 2.48 & 4.76 & 2322 \\
         & 4$'$E & 2.40 & 0.96 & 2372 \\
\\
F533-3 & 2$'$W & 2.33 & 1.45 & 12681 \\
         & C     & 1.88 & 2.01 & 12655 \\
         & 2$'$E & 2.94 & $\le$1.09 &  \\ 
\\
PGC~71626 & 4$'$W & 2.56 & 0.72 & 10037 \\
          & 2$'$W & 2.38 & 1.70 & 10013 \\
          & C     & 2.51 & 2.51 & 10018 \\
          & 2$'$E & 2.32 & 1.78 & 10015 \\
          & 4$'$E & 3.22 & $\le$0.39 &  \\
\vspace{-2 mm} \\
\hline

\end{tabular}
}
\label{}

{\scriptsize $^{a}$Three out of four scans of UGC~1752
 at the 2$'$E position were obtained with a 
band-limiting filter not optimized for the redshifted frequency of the
source. As a result, the observed integrated
flux at this position is diminished due to
filter response deterioration near the edge of the filter.}

\normalsize
\end{table}

In our present program
several of our targets have recessional velocities $V_{r}\ge$12\,000~\kms\
and hence were observed
at frequencies where 
calibration reliability and consistency at \nan\ and other radio
telescopes are less well established. 
To estimate the comparative accuracy of our flux
density calibration 
at these lower frequencies as well as recheck frequency dependent changes in
the noise diode temperature,
we examined continuum calibration data obtained at 
1400, 1425, and 1280~MHz,
from several periods over the course of the months during which 
our spectral line data were acquired (L. Alsac, private
comm.; see also Thuan et al. 2000). Over this frequency range we found
the noise diode temperature to vary by less than 10\%. Our data were
corrected for this effect based on a linear correction curve derived from
the continuum data. These calibration data also
confirm the expected internal calibration accuracy of our data
is $\sim\pm$15\% near 1420-1425MHz, but only $\sim\pm$25\% near 1280MHz.

An additional step was required for accurate flux calibration of our
\nan\ data, as
it has been found that changes have occurred in the output power of the 
calibration diode used at \nan\ since the early 1990's
(see Figure~4 of Theureau et al. 1998; see also Thuan et al. 2000), 
resulting in an overall
shift of the absolute calibration scale. This makes it
necessary to appropriately renormalize the fluxes determined
via the standard calibration techniques described above
(e.g., Theureau et al. 1998; Matthews et al. 1998; Thuan et
al. 2000). 

Matthews et al. (1998) showed via a statistical comparison
of integrated fluxes measured for $\sim$30 galaxies
at \nan\ and elsewhere that applying a
scaling factor of 1.26 to the \nan\ flux densities  very effectively corrects
for the above effect, and restores the correct normalization of the \nan\
flux scale. Matthews \& van Driel (2000) subsequently found that the
application of this same factor minimized scatter between fluxes determined
for a second sample of galaxies observed at both \nan\ and at
Arecibo. Theureau et al. (1998 and priv. comm.) also derived similar
corrections via independent observations of line calibration
sources. As a final calibration step 
we therefore apply a renormalization factor of 1.26 to all fluxes
reported in the present work.

%----- Figure 2 ------------------------------------------------------
%
\begin{figure*}[!t]
\hfil \epsfxsize 15cm \epsfbox{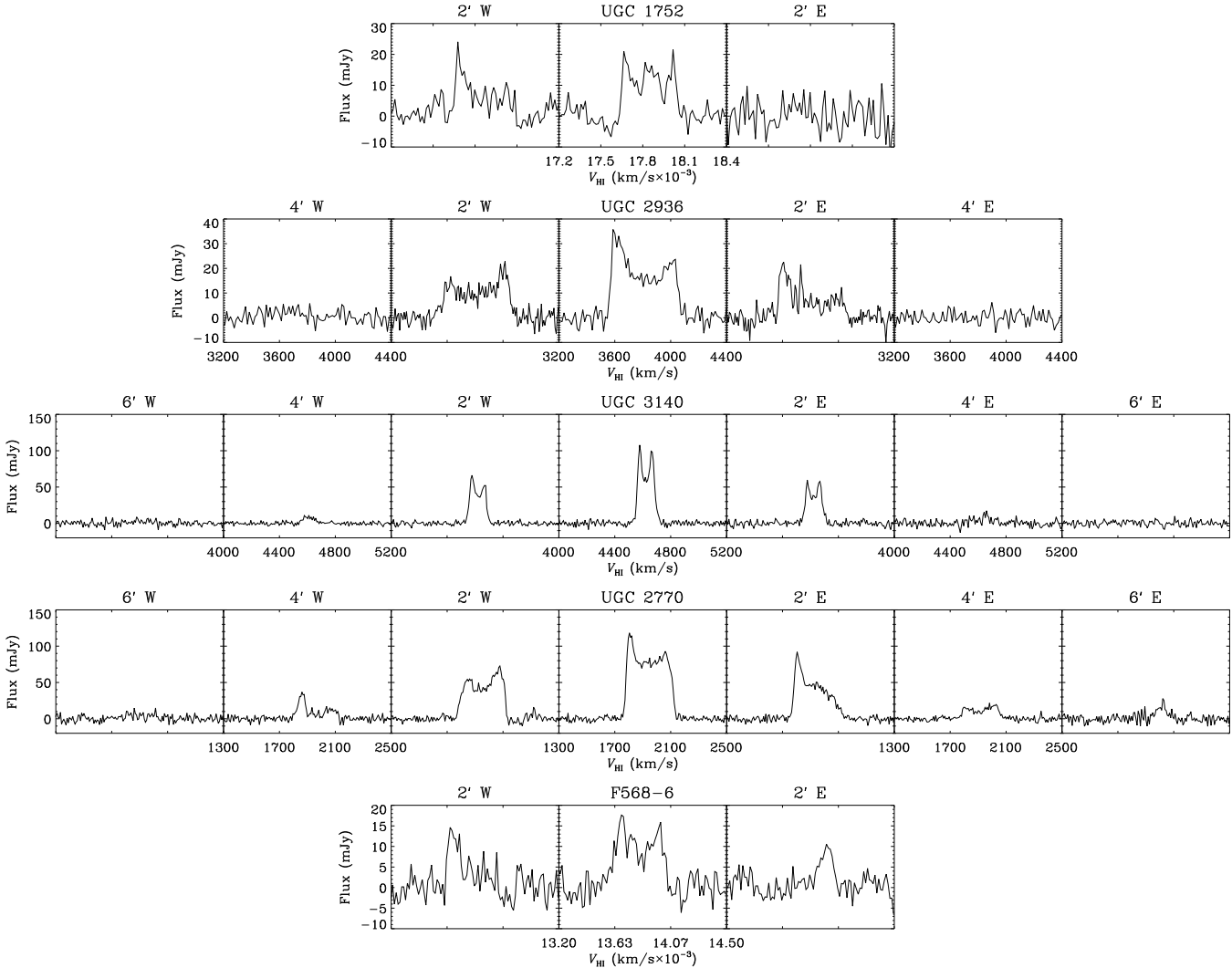}\hfil
\caption []{\nan\ 21-cm line pencil beam maps obtained for 10 LSB Giant
galaxies in our sample. Each map was obtained along an E-W line, centered at
the optical center of the galaxy. The panels for each galaxy show the
spectra obtained at positions offset by multiples of one half beamwidth
($\sim$2$'$) from the galaxy center. Axes are flux density, in
millijanskys, and velocity in \kms.}
\end{figure*}
%----------------------------------------------------------------------------
%
%----- Figure 2b ------------------------------------------------------
%
\setcounter{figure}{1}
\begin{figure*}[!t]
\addtocounter{figure}{0}
\hfil \epsfxsize 15cm \epsfbox{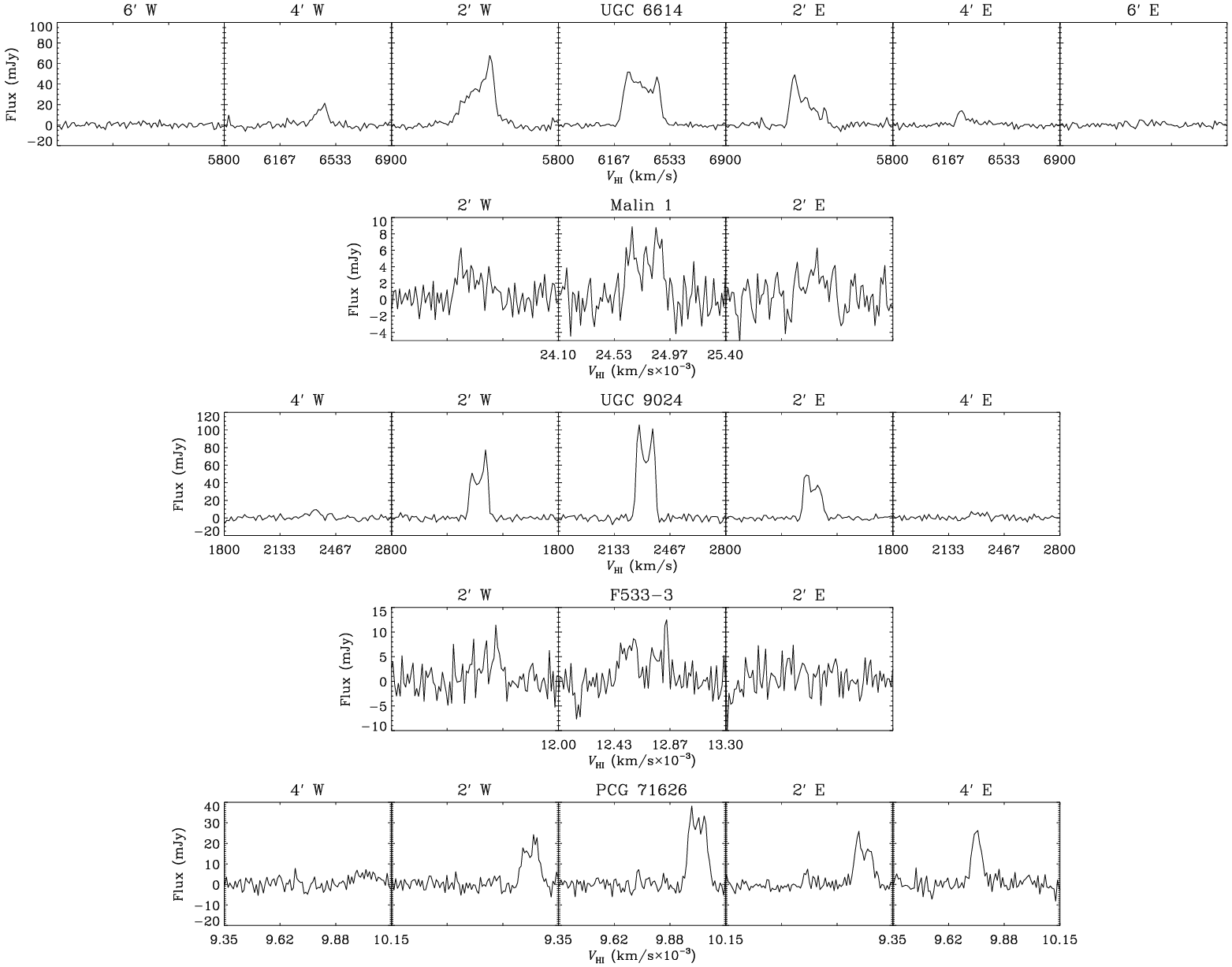}\hfil
\caption []{cont.}
\end{figure*}
%----------------------------------------------------------------------------

\subsection{Data reduction}
We reduced our \HI\ spectra using the standard DAC and SIR \nan\ spectral line 
reduction packages available at the \nan\ site. With this software 
we subtracted baselines (generally third order polynomials), averaged 
the two receiver polarizations, and applied a declination-dependent
conversion factor to convert from  units of $T_{sys}$ to flux density 
in mJy.  Because of the broad width of the lines, careful attention
was paid to baseline fitting, and scans with extremely curved
baselines were discarded. In addition, the reductions were performed
independently by two of us to check the consistency of the results.

%=====================================================================
% -----------------------------------------------------------------
%
% --- Table 4
%
\begin{table*}%
\caption{Derived \HI\ parameters} 
\begin{footnotesize}
\begin{tabular}{lccccccc}
\hline
\vspace{-2 mm} \\
 Galaxy Name & $W_{20,c}$ & $W_{50,c}$ & $V_{LSR}$ & $D$ & log$M_{HI}$ &
$M_{HI}/L_{B}$ & $D_{HI}$  \\
\vspace{-2 mm} \\

  & \kms\ & \kms\ & \kms\ & Mpc & \msun\ &  & $'$  \\

~~~~~~~~(1) & (2) & (3) & (4) & (5) & (6) & (7) & (8) \\
\vspace{-2 mm} \\
\hline
\vspace{-2 mm} \\
UGC~568    & 385 & 286 & 13579 & 181.0 & 10.25 & 0.30 &     \\

UGC~1752*  & 398 & 364 & 18072 & 241.0 & 10.85 & 1.29 & 2.0 \\

PGC~135657 & 121 &  94 & 13204 & 176.0 & 10.20 & 0.47 &     \\

UGC~2936*  & 488 & 461 &  3860 &  51.5 &  9.85 & 1.00 & 3.3 \\

UGC~3140*  & 143 & 118 &  4633 &  61.8 & 10.08 & 0.45 & 3.0 \\

NGC~2770*  & 343 & 319 &  1936 &  25.8 &  9.76 & 0.85 & 4.1 \\

PGC~135754 & 200 & 182 & 21588 & 287.8 & 10.06 & 0.27 &    \\

F568-6*    & 394 & 374 & 13738 & 183.2 & 10.52 & 0.41 & 1.3:\\

UGC~6614*  & 277 & 248 &  6245 &  83.3 & 10.42 & 1.44 & 6.0 \\

Malin-1*   & 308 & 269 & 24679 & 329.0 & 10.66 & 0.17 & 2.5 \\

PGC~45080  & 296 & 267 & 12185 & 162.5 &  9.99 & 1.72 &     \\

UGC~9024*  & 126 & 117 &  2296 &  30.6 &  9.40 & 0.92 & 2.7  \\

F530-1     & 467 & 460 & 14596 & 194.6 & 10.27 & 1.07 &      \\

F533-3*    & 375 & 367 & 12941 & 172.5 & 10.24 & 0.75 & 2.5: \\

NGC~7589  & 395: & 345: & 9056 & 120.7 & 10.01 & 0.55 &      \\

PGC~71626* & 115 &  83 & 10193 & 135.9 & 10.23 & 0.34 & 2.6  \\
\vspace{-2 mm} \\
\hline
\vspace{-2 mm} \\
\multicolumn{8}{l}{*\, indicates mapped galaxy (see Fig.~2 and Sect.~3.5).}
\end{tabular}
\end{footnotesize}
% \end{center}

\end{table*}
%======================================================================

\subsection{Measurement of \HIit\ parameters from global 
spectra}
Radial velocities, $V_{HI}$, peak flux densities, integrated line fluxes, 
velocity widths at 50\% and 20\% of peak maximum ($W_{50}$ and $W_{20}$), 
and rms noise levels of our program spectra were measured 
using our own IDL software. 
 Velocity widths were
measured interactively, by moving the cursor outward from
the profile center. Radial
velocities were defined to be the centroid of the two 20\% peak maximum
points on the profile and are quoted using the optical convention.

\subsection{Analysis of mapped galaxies}
To construct the global \HI\ profiles for each of the mapped galaxies,
we employed the procedure of Matthews et al. (1998). A
Gaussian model with appropriate sidelobes for the \nan\ beam  was
assumed (see
Guibert 1973).
We treated the beam as infinite in the N-S direction,
thus reducing the analysis to a one-dimensional problem. With our
model beam, the model galaxy flux
distributions were then iteratively integrated numerically
until the best-fit model that reproduced the
observed flux distribution in each of the telescope pointings was found. 

In all
cases  an asymmetric Gaussian \HI\ distribution (i.e., a 
lopsided Gaussian with a different $\sigma$ on the E and W sides, but
uniform height) was assumed for the \HI\ distribution of
the galaxy.
Because all of our sample galaxies were
only coarsely resolved by the \nan\ beam in the E-W direction,
use of models for the \HI\ distribution more complex than a Gaussian
(e.g., containing central \HI\ depressions,
etc.) was not attempted (see also Fouqu\'e 1984).  Moreover, we found
the simple Gaussian models produced a good match to the data in all
but two cases (F568-6 \& F533-3; see Sect.~\ref{disc}).

\subsection{Global \HIit\ spectra and measured \HIit\ parameters}
Our reduced \nan\ global \HI\ spectra for all of our target 
galaxies are shown in Fig.~1. For the
mapped galaxies, the spectra at each individual pointing are 
shown in Fig.~2. 

Parameters for the final global spectra for all of our targets,
including the mapped galaxies, are given in Table~2.
The columns in Table~2 are defined as follows:

\noindent {\bf (1)} Galaxy name.
\smallskip

\noindent {\bf (2)} Spectrum rms, in millijanskys.
\smallskip

\noindent {\bf (3)} Peak flux density
of the line profile, in millijanskys.
\smallskip

\noindent {\bf (4) \& (5)}  Raw, measured full width at 20\% and 50\% 
of the maximum
profile height, respectively, in \kms.
No correction has been applied to the raw linewidths for cosmological
stretching, instrumental resolution, or for the errors arising from
describing equal frequency-width channels by a constant velocity
width across the entire bandwidth of the spectrum (but see Table~4). 
The latter effect
is inherent in the \nan\ software, but is negligible $\la$1.5~\kms)
compared with our measurement uncertainties.
\smallskip

\noindent {\bf (6)} Heliocentric radial velocity, in \kms,
quoted using the optical convention, $V_{HI}=c(\nu_{0}-\nu)/\nu$.
\smallskip
   
\noindent {\bf (7)} Uncertainty in the heliocentric radial velocity,
in \kms\, computed following the prescription of Fouqu\'e et al. (1990).
Errors in the measured
linewidths may be estimated as $\sigma(W_{20})\approx 3\sigma(V)$ and
$\sigma(W_{50})\approx 2\sigma(V)$ (Fouqu\'e et al. 1990).
\smallskip

\noindent {\bf (8)} Raw, integrated \HI\ line flux, in \jks.  
No corrections have been applied for beam attenuation.
\smallskip

\noindent {\bf (9)} 
Uncertainty in the integrated line flux, in \jks, computed following
Fouqu\'e et al. (1990).
\smallskip

\noindent {\bf (10)} Signal-to-noise ratio of the detected line, 
defined as the ratio of the peak flux density to the spectrum rms.
\smallskip

\noindent {\bf (11)} Comments. For more detailed comments on individual
spectra, see Section~\ref{disc}.

Table~3 summarizes the raw, integrated line profile fluxes ($S_{HI}$)
and velocity
centroids ($V_{c}$) for each  pointing in our mapping observations.
In cases where no flux was detected at a
particular pointing, an upper limit to the integrated flux
was derived by multiplying the rms
noise of the spectrum by the linewidth at 50\% peak maximum from the
previous pointing. 

\bigskip
In Table~4 we tabulate several additional parameters for our target
galaxies. Columns in Table~4 are as follows:

\noindent {\bf (1)} Galaxy name
\smallskip

\noindent {\bf (2) \& (3)} $W_{20}$ and $W_{50}$ values
corrected for cosmological stretching and
spectral resolution, using the relation
\begin{equation}
W_{w,cor}=\left[W_{w,raw} + \delta_{c}\right]/(1+z)
\end{equation}
\noindent (see Haynes \& Giovanelli 1984).
Here $W_{w,raw}$ is the raw observed linewidth at $w$=20\%
or $w$=50\% peak maximum, $z=V_{HI}/c$, and $\delta_{c}$ is given by
\begin{equation}
\delta_{c}=(0.014\omega - 0.83)\delta_{R}
\end{equation}
\noindent where $\omega$=20 and $\omega$=50 for $W_{20}$ and $W_{50}$,
respectively, and $\delta_{R}$ is the velocity resolution of the
measured spectrum (see Bottinelli et al. 1990).
No corrections were applied
for inclination angle of the source or for turbulent
motions.
\smallskip

\noindent {\bf (4)} Radial velocity, in \kms, corrected to the Local
Standard of Rest, following the prescription of Sandage \& Tammann (1981):
\begin{equation}
V_{LSR}=V_{HI}-79{\rm cos}l{\rm cos}b+296{\rm sin}l{\rm cos}b-36{\rm
sin}b \kms.
\end{equation}

\smallskip

\noindent {\bf (5)} Galaxy distance in Mpc, computed from $D=V_{LSR}/H_{0}$.
  \smallskip
 
\noindent {\bf (6)} Logarithm of the \HI\ mass in solar units, computed  from
the integrated line flux $S_{HI}$ 
in column~8 of Table~2 and using the relation
$M_{HI}=2.36\times10^{5}D^{2}S_{HI}$. 
\smallskip

\noindent {\bf (7)} Ratio of the \MHI\ mass to the optical $B$-band
luminosity $L_{B}$, in solar units. 
$L_{B}$ was derived from the mean of the absolute
$B$ magnitudes for each galaxy given in Table~6 (discussed below) and
assuming a solar absolute magnitude of $M_{B,\odot}=$5.48.
For NGC~7589 a
$B$-band magnitude was taken from the NED database.

\noindent {\bf (8)} Rough estimate of the \HI\ diameter of the source in
arcminutes. Estimates were made only for mapped galaxies where flux
was detected at 2 or more positions (see Table~3). 
Following Fouqu\'e (1984), we
define the \HI\ diameter as the isophote enclosing half of the \HI\
mass in a flat \HI\ disk model, which for a Gaussian \HI\ surface density,
is equal to the FWHM of the model (Fouqu\'e 1984).  Because of
the elongation of the \nan\ beam and the fact that our maps were
obtained along an E-W axis, a correction to the raw \HI\ diameter
for the position angle and
inclination of the source was also applied. Hence,
\begin{equation}
D_{HI}=Q^{-1}D^{EW}_{HI}
\end{equation}
\noindent where:
\begin{equation}
Q^{2}={\rm sin}^{2}(PA) + R^{-2}_{H}{\rm cos}^{2}(PA).
\end{equation}
\noindent
Here, $R_{H}$ is the ratio of the major to the minor axis of
the \HI\ distribution and $PA$ is the position angle of the major
axis. We assume $R_H$=$(a/b)_{optical}$ and adopt the photometric
position angle for $PA$ (Fouqu\'e 1984). Typical errors for this
method of estimating $D_{HI}$ are $\sim 30\pm30$\% (see Fouqu\'e 1984).

\section{Discussion\protect\label{disc}}
%
% -----------------------------------------------------------------
%
% --- Table 5: HI data - literature values 
%
\begin{table*}
\caption{Comparison of published \HI\ line data with new measurements}
{\scriptsize
\begin{tabular}{lrrrrrl|lrrrrrl}
\hline
\vspace{-2 mm} \\
Ident. & $V_{HI}$ & $W_{50,c}$ & $W_{20,c}$ & $S_{HI}$ & Tel. & Ref. & Ident. & 
 $V_{HI}$ & $W_{50,c}$ & $W_{20,c}$ & $S_{HI}$ & Tel. & Ref.\\ 
  & km/s  & km/s & km/s & Jy km/s & & &   & km/s  & km/s & km/s & Jy km/s & & \\
\vspace{-2 mm} \\
\hline
\vspace{-2 mm} \\
UGC 1752   & 17836 & 384 &     &  3.93 & A & GH89 & Malin-1   & 24750 &     & 341 &  3.5  & A & Bo87 \\ 
           & 17861 & 364 & 398 &  4.97 & N & New  &           & 24745 & 315 & 340 &  4.6  & B & IBpc \\ 
           &       &     &     &       &   &      &           & 24705 & 295 & 355 &  2.5  & A & IBpc \\ 
UGC 2936   &  3809 &     & 490 &  8.36 & A & He83 &           & 24755 & & 322     &  2.5  & V & P97  \\
           &  3823 &     & 490 &  8.4  & A & HG84 &           & 24784 & 269 & 308 &  1.68 & N & New \\
           &  3816 &     & 513 & 11.7  & G & TC88 &           &       &     &     &       &   &      \\
           &  3817 & 493 & 499 &  9.0  & A & S95  & PGC 45080 & 12264 &     &     &       & A & S95 \\ 
           &  3811 & 484 &     & 13.29 & A & H99  &           & 12339 & 267 & 296 & 1.53  & N & New \\ 
           &  3813 &     & 500 & 13.6  & V &  P99 &           &       &     &     &       &   &      \\
           &  3811 & 461 & 488 & 11.20 & N &  New & UGC 9024  &  2285 &     &     &       &   & Bo85b  \\ 
           &       &     &     &       &   &      &           &  2326 & 114 & 128 &  8.7  & A & L87   \\  
UGC 3140   &  4650 &     &     &       & G & S78  &           &  2323 & 117 &     &  9.3  & A & H97    \\
           &  4630 & 161 & 228 & 12.08 & N & Bt82 &           &  2323 & 115 & 131 &  8.4  & N & T98   \\ 
           &  4622 & 125 & 148 & 12.30 & J & SD88 &           &  2321 & 117 & 126 & 11.38 & N & New   \\
           &  4623 & 118 & 143 & 13.35 & N & New  &           &       &     &     &       &   &      \\ 
           &       &     &     &       &   &      & F530-1    & 14339 & 488 &     &  1.7  & A & S92   \\ 
NGC 2770   &  1953 & 330 & 347 & 33.26 & G & FT81 &           & 14335 & 460 & 467 & 2.01  & N &  New \\ 
           &  1950 & 334 & 357 & 37.84 & J & SD88 &           &       &     &     &       &   &     \\    
           &  1944 & 327 &     & 24.10 & A & C93  & F533-3    & 12660 & 406 & 430 &  2.5  & A & GH89  \\
           &  1948 & 340 & 373 & 33.88 & W & RA96 &           & 12659 & 398 &     &  2.2  & A & S92    \\ 
           &       & 324 &     & 35.32 & A & H97  &           & 12670 & 367 & 375 &  2.41 & N & New   \\ 
           &  1947 & 319 & 343 & 36.90 & N & New  &          &       &     &     &       &   &     \\ 
           &       &     &     &       &   &      & NGC 7589  &  8938 & 356 & 363 &       & A & S95  \\  
F568-6     & 13827 & 386 &     &  3.48 & A & S92  &           &  8925 &     & 402 &  2.7  & V & P97   \\ 
           & 13835 &     & 392 &  4.4  & V & P97  &           &  8866 & 345: & 395: & 2.90 & N & New   \\ 
           & 13829 & 374 & 394 &  4.10 & N & New  &           &       &     &     &       &   &     \\ 
           &       &     &     &       &   &      & PGC 71626 & 10022 &  91 & 118 &  2.6  & N & T98  \\   
UGC 6614   &  6351 & 244 & 278 &  7.08 & A & Bo85a &          & 10016 &  83 & 115 & 3.86  & N & New   \\ 
           &  6358 & 236 & 286 &  9.56 & A & HR89  &          &       &     &     &       &   &     \\ 
           &  6353 &     & 287 & 15.0  & V & P97  &           &       &     &     &       &   &      \\ 
           &  6351 & 248 & 277 & 15.79 & N & New  &           &       &     &     &       &   &      \\ 
\vspace{-2 mm} \\
\hline
\end{tabular}
\begin{tabular}{llllll}
\vspace{-2 mm} \\
\multicolumn{6}{l}{References:} \\
\vspace{-2 mm} \\
Bo85a & Bothun  et al. (1985a)          &
Bo85b & Bothun et al. (1985b)           &
Bo87  & Bothun et al. (1987)            \\
Bt82  & Bottinelli  et al. (1982)       &
C93   & Chengalur et al. (1993)         &
FT81  & Fisher \& Tully (1981)          \\
GH89  & Giovanelli \& Haynes (1989)     &
HG84  & Haynes \& Giovanelli (1984)     &
H97   & Haynes  et al. (1997)           \\
H99   & Haynes  et al. (1999)           &
He83  & Hewitt et al. (1983)            &
Hu83  & Huchra et al. (1983)            \\
HR89  & Huchtmeier \& Richter (1989)      &
IBpc  & Impey \& Bothun (1989), see HR89  &
L87   & Lewis  (1987)                    \\
New   & This work                       &
P97   & Pickering et al. (1997)         &
P99   & Pickering et al. (1999)         \\
RA96  & Rhee \& van Albada (1996)       &
S92   & Schombert et al. (1992)         &
S78   & Shostak      (1978)             \\
S95   & Sprayberry et al. (1995b)       &
SD88  & Staveley-Smith \& Davies (1988) &
T98   & Theureau et al. (1998)          \\
TC88  & Tifft \& Cocke (1988)           & 
      &                                 \\
\vspace{-2 mm} \\
\multicolumn{6}{l}{Telescope codes:} \\
\vspace{-2 mm} \\
A\, & Arecibo 305m       & 
B \,& Green Bank 43m     &
E\, & Effelsberg 100m    \\
G\, &  Green Bank 91m    &
J\, & Jodrell Bank 64m   & 
N\, & \nan\ 94m equiv    \\
V   & VLA                &
W\, & Westerbork         & 
    &                    \\
\vspace{-2 mm} \\
\hline
\end{tabular}
}
\label{hilit}
\normalsize
\end{table*}
% -----------------------------------------------------------------

\subsection{The \HIit\ properties of LSB Giants}
Not surprisingly, all of the LSB Giant galaxies observed in the present
program are found to be 
very \HI-rich systems, having \mhi\ values ranging from
$2.5\times 10^{9}$~\msun--7.1$\times 10^{10}$~\msun (Table~4). 
The bulk of the sample
galaxies are not only \HI\ rich in an absolute sense, but are
also \HI-rich both for their Hubble types and for their
optical luminosities---six of the galaxies in the present LSB Giant
sample have $M_{HI}/L_{B}\ga$1 (in solar units;
see Table~4). More typical values of
$M_{HI}/L_{B}$ for
high surface brightness spirals are: $\sim$0.2 (type Sab-Sb);
$\sim$0.3 (type Sbc-Sc); $\sim$0.4 (type Scd-Sd) (Roberts \& Haynes 1994).

As expected for massive, \HI-rich disks, broad \HI\ linewidths are also
hallmarks of LSB Giants. Because a number of our program galaxies
have rather low inclinations ($i\le 40^{\circ}$), it is difficult to estimate
the true rotational velocities in some cases, but even many of the 
uncorrected linewidths tend to be quite broad (e.g., $W_{20,c}=$398~\kms\
for UGC~1752, which is seen nearly face-on). Only PGC~71626 seems
to have a surprisingly narrow linewidth ($W_{20,c}$=115~\kms) given its
inclination ($i=61^{\circ}$) and Hubble type (SBb or SBc).

The majority of the LSB Giants in our study exhibit relatively normal, classic,
double-horned rotational profiles (Fig.~1). However, we do see a few
unusual cases as well;  UGC~568 
shows a peculiar, strongly asymmetric
profile; PGC~45080 and NGC~7589 both have  rather squarish global profiles 
with no clear
rotation horns; and finally,
F530-1 has a double-horned profile that is strongly ``cleft''
in the middle. PGC~135754 may also have a similar cleft morphology, but
our detection of this source is relatively weak.
Among the 11 relatively normal, double-horned profiles, 3 cases show modest
asymmetries (UGC~2936, NGC~2770, and F568-6). Thus we find no evidence
that strong \HI\ asymmetries are a general feature of LSB Giants. 
We comment further on our individual
spectra below.

\subsection{Comparison of new \HIit\ data with previous measurements
and comments on individual objects}

In order to compare the \HI\ parameters derived in the present
work with past measurements,
we have compiled global \HI\ parameters from the literature
for the 13 of our sample galaxies which have previously been observed
in the 21-cm line. Table~5 summarizes previously measured recessional 
velocities,
linewidths, and integrated
line fluxes for each case, and indicates the telescope at which the data
were obtained, as well as the reference for the quoted parameters.

An examination of Table~5 shows that for LSB Giants observed
by more than one group, significant discrepancies frequently
exist between
the measured parameters, particularly the global integrated line fluxes.
We discuss each of the 16 targets observed in the present
work in more detail below, including comments on these
discrepancies.
As part of our investigation of  our LSB Giant targets, we also searched 
the vicinity of  each of the 16 objects (i.e., an area about 1.5 
times the \nan\
HPBW) for nearby galaxies 
which could possibly give rise to confusion in our observed \HI\ profiles.
 For this we employed the NED and LEDA databases, 
as well as optical images extracted 
from the Digitized Sky Survey (DSS). Results of these searches are also
described below.

\smallskip

\noindent {\bf UGC 568}: no \HI\ data have been previously published 
for this object. Our global spectrum appears quite peculiar, having a
strong asymmetry, with a broad, single ``horn'' on the low velocity
side of the profile.
Published optical redshifts show a discrepancy of over 400~\kms 
($V_{opt}$=13\,600~\kms, Melnick \& Quintana 1981;  
$V_{opt}$=13\,198$\pm$275~\kms,
Impey et al. 1996). Our new \HI\ velocity corresponds to the mean
of the optical redshifts.
Several other galaxies were found within our search area, 
but they do not seem candidates for an \HI\ confusion that could
explain our peculiar global \HI\ profile:
CGCG~384-029 is a $B_T$=15.0 mag Sbc spiral with an optical velocity 
886~\kms\ higher than the LSB Giant (Dale et al. 1998; 
Katgert et al. 1998), while PGC~3280 is an S0 with a recessional velocity
1524~\kms\ lower than the LSB; MCGG~+00-03-027
is a 15.5 mag galaxy without known 
velocity, but it is classified as S0/a (LEDA) or Sa (NED), and is therefore 
expected to be gas-poor. 

\smallskip

\noindent {\bf UGC 1752}: Our new total integrated \HI\ flux
($S_{HI}=$5.13~\jks) is somewhat larger than that measured at Arecibo
by Giovanelli \& Haynes (1989; $S_{HI}=3.93$~\jks), suggesting this
galaxy has extended \HI\ emission.
\smallskip

\noindent {\bf PGC 135657}: no \HI\ data have been 
previously published for this object. Our \HI\ detection has a
recessional velocity 407~\kms\ larger than the
optical velocity of
$V_{opt}$=12\,701$\pm$275 \kms\ reported for PGC~135657 by Impey et al. (1996).
In the present survey we detect additional marginal features at the
edges of our bandpass, near $V_{HI}=$12\,100~\kms\ and
$V_{HI}=$13\,300~\kms. There are a number of
small angular size galaxies in this field, but none appear as obvious
candidates for interlopers.
\smallskip

\noindent {\bf UGC 2936}: The Arecibo spectrum of UGC~2936
published by Sprayberry (1995b) is highly asymmetric; however only
a modest asymmetry is seen in our new data and in the VLA spectrum
of Pickering et al. (1999). Our new integrated line flux of 11.23~\jks\
agrees to within errors with that of Tifft \& Cocke (1988) ($\sim$11.7~\jks). 
In comparison,
the Arecibo spectra by 
Hewitt et al. (1983),
Haynes \& Giovanelli (1984), and Sprayberry et al. (1995b)
all  yield considerable lower integrated line fluxes 
($\sim$8.4-9.0~\jks), while
our new integrated flux value is somewhat lower than
the Arecibo measurement  of Haynes et al. (1999 13.29~\jks) and the 
VLA measurement
of Pickering et al. (1999; 13.6~\jks). 

CO(1-0) and CO(2-1) line emission were detected from UGC~2936  at SEST 
and IRAM by 
Chini et al. (1996), making this one of the rare galaxies classified
as an LSB Giant that has been detected in CO (see also Knezek 1993,1998).
However, Schmelz et al. (1986) failed to detect 
OH line absorption from the object.
\smallskip

\noindent {\bf UGC 3140 (= NGC 1642)}: This galaxy forms a wide pair 
(\am{8}{3} projected separation; 
$\Delta V_{opt}$=332$\pm$85 \kms) with 15.0 mag Sb spiral UGC~3141.
Their E-W separation is \am{3}{3} and our \nan\ spectrum does not
appear to be confused. Our new linewidth measurements
are in good agreement with the Jodrell Bank measurements
of Staveley-Smith \& Davies (1988) and measurements we performed 
from new HIPASS data (see Barnes et al., in prep. for a description of
HIPASS), but are lower than the values
published by Bottinelli et al. (1982). No published \HI\ observations are
available for UGC~3141.
\smallskip

\noindent{\bf NGC 2770}: Short Westerbork radio synthesis 21-cm line 
observations
(Rhee \& van Albada 1996; Broeils \& Rhee 1997) indicate 
that the \HI\ distribution of NGC~2770 
at a surface density level of 1 \Msun\ pc$^{-2}$ is about 
1.7 times as large as the optical ($D_{25}$) diameter and that its 
mean \HI\ surface density
averaged over the entire \HI\ disk $<$$\sigma_{HI}$$>$ 
is 4.5 \Msun\ pc$^{-2}$, 
which is higher than typical values for  an Sc. Our new spectra
show a slightly higher peak flux density on the low-velocity side of the global
line profile, hinting at a possible weak asymmetry not seen in the 
Westerbork observations.
\smallskip

\noindent {\bf PGC 135754}: no \HI\ data have been previously
published for this object and it
was only weakly detected in the present study. Our new \HI\ velocity
is 434~\kms\ higher than the optical value reported by Impey
et al. (1996; 21335$\pm$275~\kms). 
\smallskip

\noindent {\bf F568-6 (= Malin-2 = PGC 86622)}:  
\HI\ flux appears to have been 
missed in the Arecibo observations of F568-6 by Schombert et al. (1992). Our
new value of $S_{HI}$=4.20~\jks\ shows good agreement with
the value derived from the VLA observations of 
Pickering et al. (1997; 4.4~\jks).

\smallskip

\noindent {\bf UGC 6614}: Our new integrated \HI\ flux
of 15.94~\jks\ is significantly larger than previous
measurements from Arecibo (Bothun et al.
1985a; Haynes \& Giovanelli 1989). However, Pickering et al. (1997)  measured
$S_{HI}$=15.0~\jks\ from VLA data, consistent with our results.  
Our \HI\ diameter
estimate for this object is very poorly constrained, as we find
a Gaussian \HI\ model gives a poor
fit to the data. This is reaffirmed by the VLA data of Pickering et al. (1997),
which show a signficant central depression in the \HI\ distribution of this
galaxy.
\smallskip
%other \HI\ line imaging : van der Hulst et al. 1993
%One galaxy was found inside the \nan\ beam area,
%KUG~1136+173, a 15.7 mag spiral of \am{0}{6} diameter
%and of unknown redshift, \am{5}{6} due south of the
%LSB.

% Checked: Schommer \& Bothun (1983) - HI detection, probably Arecibo;
%          only HI mass quoted, distance (not given) from Bothun's Thesis
% Checked: Martin (1998) HI imaging catalog - not listed

\noindent {\bf Malin-1 (= PGC 42102)}: One published Arecibo observation 
of Malin-1 (see Table~5)
gives an integrated \HI\ line flux (2.5~\jks; Bothun et al. 1987)
comparable to the VLA value derived by
Pickering et al. (1997),
while a re-reduction of the same Arecibo 
data yielded $S_{HI}$=3.5 Jy \kms\ (Impey
\& Bothun 1989); finally  Impey \& Bothun (1989) measure
$S_{HI}$=4.6~\jks\ from a Green Bank 43m spectrum. Our new data
indicate that the Arecibo and Green Bank
fluxes previously reported by Impey \& Bothun (1989)
are likely to be significantly overestimated. In contrast
our new integrated flux (1.80~\jks) is somewhat lower than the VLA measurement
of Pickering et al. (1997), although our values marginally agree to
within errors, since at the redshifted frequency of Malin-1 ($\sim$1312~MHz),
our  absolute 
calibration uncertainties are expected to be as high as $\sim\pm$25\%
(see Section~\ref{cal}). In terms of spectral morphology, our new
global spectrum shows excellent agreement with that derived by
Pickering et al.

No CO(1-0) line emission was detected in the central regions of Malin-1
by Radford (1992), with an estimated upper limit of 4.7$\times 10^9$ \Msun\
to the H$_2$ mass, and Braine et al. (2000) recently reported an upper limit of
$M_{H2}/M_{HI}<0.03$. 

\smallskip

\noindent {\bf PGC 45080}: Our new integrated line flux is 
smaller than the values reported by Sprayberry et
al. (1995b), and our new spectrum lacks the distinct rotation horns
seen in the Sprayberry et al. data. Given the small optical size of this object
($D_{25}$=\am{0}{79}) it seems unlikely significant flux extended outside our
beam. There are also no obvious interlopers in this field that
would have caused confusion in the Sprayberry et al. spectrum, hence
the discrepencies in the two spectra are probably due to the lower
signal-to-noise of our present data.
\smallskip

\noindent {\bf UGC 9024}: Our new integrated flux value of 11.41~\jks\
is higher than past values measured at Arecibo or with single-pointing
\nan\ measurements (Bothun et al. 1985; Lewis 1987; Haynes et
al. 1997; Theureau et al. 1998), suggesting this galaxy has extended
\HI\ emission.
\smallskip

\noindent {\bf F530-1 (= PGC 87136)}: We detect only a slightly higher
integrated flux (2.07~\jks) than the Arecibo value of
Schombert et al. (1992; 1.7~\jks).  Our spectrum shows
an unusual ``cleft'' morphology, although our signal-to-noise is only
modest due to the broad linewidth of this source.
\smallskip

\noindent{\bf F533-3 (= PGC 68495)}: 
No flux is clearly detected at 2$'$E in our mapping observations of
this galaxy; this may be due to the
broadness and relative weakness
of the line profile and/or an \HI\ distribution that is very lopsided
relative to the optical center of the galaxy. 
As a result, our \HI\ diameter estimate
for F533-3
very uncertain. 
\smallskip 

\noindent{\bf NGC 7589}: NGC~7589 was
mapped in \HI\ at the VLA by Pickering et al. (1997). Our new \HI\
profile hints at a possible small amount of additional flux on the
low-velocity side of the profile compared with the VLA global
spectrum; this is probably caused by confusion with F893-29, which was also
detected in \HI\ at the VLA. F893-29 is  a 17.1 mag
Sb? spiral (NED) or dwarf (Pickering et al. 1997), 
\am{3}{3} SW of the LSB Giant, with a E-W separation
of \am{2}{2}.
F893-29 has a systemic \HI\ velocity of  8768 \kms\ (Pickering et
al. 1997), consistent
with the velocity of the ``extra'' flux we observe associated with our
global profile of NGC~7589. Our new integrated line flux (2.96~\jks)
agrees well with that reported by Pickering et al (1997; 2.7~\jks),
hence the flux contamination from this second source appears to be
minimal. However our line profile widths appear to be overestimated by
$\sim$100~\kms\ due to this second source, 
hence corrections have been applied to the $W_{20}$
and $W_{50}$ measurements in Tables~2 \& 4.
%W$_{50}$$\sim$170 \kms; its integrated line flux is 1.6 Jy \kms.
\smallskip

%========================================================================
% -----------------------------------------------------------------
%
% --- Table 6: Optical data
%
{\scriptsize
\begin{table*}
\caption{Optical data from the literature}
\begin{tabular}{lcccccl}
% \smallskip \\
\hline
\vspace{-2 mm} \\
Ident. & $V_{opt}$ & $B_{tot}$ & $\mu(0)$
   & $h_{r}$ & $M$ & Ref. \\ 
 & km/s & mag & mag $''^{-2}$ & kpc & mag & \\
\vspace{-2 mm} \\
\hline
\vspace{-2 mm} \\
UGC 568    & 13197 & 14.76 & 23.0 & 17.1 & $-$21.53 & S95 \\
           & 13198 & 14.9  & 19.9 &      & $-$21.42 & I96 \\
UGC 1752   &       & 15.5$^{a}$ & 22.7 & 20.5 & $-$21.38 & K93   \\
PGC 135657 & 12701 & 15.47 & 22.6 & 11.1 & $-$20.74 & S95\\
           & 12701 & 15.3  & 20.9 &      & $-$20.92 & I96 \\
UGC 2936   &       & 13.97 &   22.23 & 9.24 & $-$19.58 & S95 \\
           &  4065 & 15.1  & 22.1 &   & $-$18.42 & I96\\
           &  3822 &       & 20.3$^{b}$ & 8.4 & $-$21.1$^{b}$ & P99\\
UGC 3140   &       & 13.34 & 20.9 & 3.9 & $-$20.61 & dJ96 \\
           &       & 13.4$^{a}$ & 22.8 & 9.8 & $-$20.55 & K93 \\
NGC 2770   &       & 12.94 & 20.8 & 13.0 & $-$19.09 & K85 \\
PGC 135754 & 21335 & 16.40 & 22.5 & 11.2 & $-$20.99 & S95\\
           & 21335 & 16.1  & 21.0 & & $-$21.22 & I96 \\
F568-6     &       & 14.57 & 23.2 & 21.0 & $-$21.79 & MB94  \\
           &       &       & 22.1$^{b}$ & 18.3 & $-$23.6$^{b}$ & P97 \\
UGC 6614   &       &       & 22.9$^{b}$ & 13.6 & $-$22.3$^{b}$ & P97 \\
           &       & 14.43 & 24.3 & 15.8 & $-$20.29 & MB94 \\
           &       &14.6$^{a}$ & 22.3 & 9.1 & $-$20.04 & K93 \\   
Malin-1    &       &       & 26.0$^{c}$ & 82.0 & $-$22.9$^{c}$ & P97 \\
           &       &       & 26.4       & 73.2 &               & S93 \\ 
           &       &       & 25.5$^{c}$ & 73.2 & $-$23.12      & IB89 \\
PGC 45080  &       & 17.48 & 22.8 & 6.45 & $-$18.65 & S95 \\
           & 11674 & 17.0 & 22.4 & & $-$19.12 & I96 \\
UGC 9024   &       & 14.90 & 24.1 & 4.3 & $-$17.58 & dJ96 \\
           &       & 14.74 & 24.4 & 7.4 & $-$18.46 & MB94 \\
F530-1     &       & 16.3$^{a}$ & 21.8 & 12.2 & $-$20.11 & K93  \\
F533-3     &       & 15.7$^{a}$ & 23.5 & 26.5 & $-$20.44 & K93  \\
NGC 7589   &       &       &  23.3$^{b}$ & 12.6 & $-$21.9$^{b}$ & P97\\
PGC 71626  &  9520 & 14.48 & 22.1 & 12.6 & $-$21.08 & S95\\
           &  9521 & 14.1 & 20.0 & & $-$21.42 & I96 \\
\vspace{-2 mm} \\
\hline
\vspace{-2 mm} \\
\multicolumn{7}{l}{Notes:} \\
\vspace{-2 mm} \\
\multicolumn{7}{l}{$^{a}$measurement within 26 mag arcsec$^{-2}$ isophote} \\
\multicolumn{7}{l}{$^{b}$measurement in $R$ band} \\
\multicolumn{7}{l}{$^{c}$measurement in $V$ band} \\
\vspace{-2 mm} \\
\multicolumn{7}{l}{References:} \\
\vspace{-2 mm} \\
\multicolumn{3}{l}{dJ96 de Jong (1996)}          & \multicolumn{4}{l}{I96 Impey et al. (1996)}  \\
\multicolumn{3}{l}{IB89 Impey \& Bothun (1989)}  & \multicolumn{4}{l}{K85 Kent (1985)}  \\
\multicolumn{3}{l}{K93 Knezek (1993)}            & \multicolumn{4}{l}{MB94 McGaugh \& Bothun (1994)} \\
\multicolumn{3}{l}{S93 Sprayberry et al. (1993)} & \multicolumn{4}{l}{S95 Sprayberry et al. (1995b)}  \\
\multicolumn{3}{l}{P97 Pickering et al. (1997)}  & \multicolumn{4}{l}{P99 Pickering et al. (1999)}  \\
\vspace{-2 mm} \\
\hline
\end{tabular}
\label{reldet}
\end{table*}
}
\normalsize
%====================================================================

\noindent {\bf PGC 71626}: PGC~71626 was detected at a recessional velocity
consistent with that recently reported by Theureau et al. (1998), but
somewhat offset from the published optical velocity of 9520~\kms\
reported by Sprayberry et al. (1995b) and Impey et al. (1996).
Our new integrated flux value of 3.94~\jks\ is higher than the value
of 2.6~\jks\ reported by Theureau et al. (1998) based on a
single-pointing \nan\ observation. An additional \HI\ source was
detected in our spectrum at $V_{HI}=9753\pm4$~\kms, with
$W_{20}=73$~\kms\ and $S_{HI}=1.52\pm0.48$~\jks. This may be
contamination from the S? galaxy MCG~-01-59-026 (of unknown redshift),
at a projected
distance of \am{4}{5} from PGC~71626, or due to an uncatalogued LSB
dwarf visible on the DSS.

\subsubsection{Optical parameters for the LSB Giants}
All of the LSB Giant galaxies observed in the present survey
have optical surface photometry available in the literature.
To offer a more complete picture of the nature of the objects in our
sample, we have compiled a summary of several optical
and photometric parameters for each galaxy in Table~6.

Radial velocities quoted in Table~6 are 
heliocentric values derived
from optical measurements. $B_{tot}$ values are the
total $B$-band magnitudes
derived
from extrapolated exponential disk fits unless otherwise noted.
$\mu(0)$ is the extrapolated central disk surface brightness
in the $B$-band (unless otherwise specified) and is
{\it uncorrected} for inclination and internal 
extinction. $h_{r}$ is the exponential
disk scale length in kpc. $M$ is the absolute magnitude ($B$-band,
unless otherwise noted)
derived from the total magnitude in column 3, and assuming 
$D=V_{opt}/H_{0}$ if no distance was given in the original reference;
no internal extinction corrections were applied. Corrections to the
values in Table~6 for
Galactic extinction, $k$-corrections, and
cosmological surface brightness dimming
were generally applied by the original authors.

The values given in Table~6 show that there is a great deal of scatter
in the photometric parameters reported for LSB Giant galaxies. Surface
photometry of LSB galaxies is always challenging, but it is further
complicated for LSB Giants by the fact that many of these galaxies
have significant bulges, leading to ambiguity in the bulge-disk
decompositions
and  hence vastly different derived scale
lengths and central surface brightness values from different workers.
Offsets between recessional velocities derived via optical and
\HI\ observations are also common (cf. Table~5). 

We include Table~6
partly as a caution to readers interested in understanding the global
properties of Giant LSB galaxies that there still exist large
uncertainties in the measured parameters for such objects, and to
highlight the need for additional surface photometry (and \HI\
measurements) of 
such galaxies. Such data will be necessary before we can distinguish whether
all LSB Giants such as the ones in the present study form a distinct
class of object, as suggested by Hoffman et al. (1992), or whether
some or most of these objects may form a natural and continuous extension
of other low and moderate surface brightness disk galaxies.
Well-established photometric parameters will also be key to placing
the LSB Giant galaxies on the Tully-Fisher diagram.

\acknowledgements{
LDM acknowledges the financial support of a Jansky Postdoctoral
Fellowship from the National Radio Astronomy Observatory (NRAO), which 
is a facility of the
National Science Foundation operated under cooperative agreement by
Associated Universities, Inc.
We have made use of the NASA/IPAC Extragalactic Database (NED) which is
operated by the Jet Propulsion Laboratory, California Institute of
Technology, under contract with the U.S. National Aeronautics and Space
Administration, as well as the Lyon-Meudon Extragalactic Database              
(LEDA) supplied by the LEDA team at the CRAL-Observatoire de                  
Lyon (France), and the Digitized Sky
Surveys (DSS), which were produced at the Space Telescope Science
Institute under U.S. Government grant NAG W-2166.
The Unit\'e Scientifique \nan\ of the Observatoire de Paris is associated as 
Unit\'e de Service et de Recherche USR No. B704 to the French Centre National 
de Recherche Scientifique (CNRS). \nan\ also gratefully acknowledges 
the financial support of the R\'egion Centre in France.
}

\end{document}